\documentclass{article}
\usepackage{spconf,amsmath,graphicx,hyperref,enumitem, booktabs,xcolor}
\usepackage{microtype}

\title{RESTORE, SEPARATE, RESTORE: \\A MODULAR FRAMEWORK FOR MUSIC SOURCE RESTORATION}

\name{Tobias Morocutti$^{\star}$ \qquad Emmanouil Karystinaios$^{\star}$ \qquad Gerhard Widmer$^{\star,\dagger}$
}
  
  \address{$^{\star}$Institute of Computational Perception, Johannes Kepler University Linz, Austria \\
      $^{\dagger}$LIT Artificial Intelligence Lab, Linz, Austria \\
      }

\begin{document}
%
\maketitle
%


\begin{abstract}
Music Source Restoration (MSR) seeks to recover original, unprocessed instrument stems from mixed, mastered, and possibly degraded recordings. Unlike conventional source separation, which treats the mixture as a linear sum of clean sources, MSR must additionally invert nonlinear production effects, such as equalization and compression, and transmission-related degradations, such as codec artifacts. We propose a three-stage framework built around this distinction: (1) a mixture restoration model that addresses degradation before separation, (2) a single model that separates the restored mixture into eight target stems (\textit{vocals}, \textit{guitars}, \textit{keyboards}, \textit{synthesizers}, \textit{bass}, \textit{drums}, \textit{percussion}, and \textit{orchestra}), and (3) stem-specific restoration experts fine-tuned on the separator's own residual artifacts. Each stage improves restoration quality over the previous one on the MSR Challenge test set. We release code and models to support future research in MSR\footnote{\href{https://github.com/theMoro/music_source_restoration}{https://github.com/theMoro/music\_source\_restoration}}.
\end{abstract}
\section{Introduction}
\label{sec:intro}

Real-world music recordings rarely correspond to simple sums of clean, unprocessed sources. Whether captured in the studio, during live performances, or preserved in historical archives, individual sources and mixtures may be affected by equalization, compression, saturation, reverberation, mixing and mastering, as well as noise, codec artifacts, or other forms of degradation. These transformations depart from the clean-source linear-mixture assumption underlying most music source separation systems~\cite{cano2019mss}. Music Source Restoration (MSR)~\cite{zang2025msr} addresses this broader setting by aiming to recover the original, unprocessed sources from an observed recording, ideally approaching the signals that could have been captured directly from the individual instruments or voices. 


MSR was first formalized in~\cite{zang2025msr}, which introduced a dataset of unprocessed music stems, RawStems. Soon after, \cite{zang2025msrbench} established the first MSR benchmark using professionally mixed rather than synthetically summed stem-mixture pairs, evaluated under different degradations, which prompted the inaugural ICASSP 2026 MSR Challenge~\cite{zang2026msrchallenge}. Most participating systems relied on BS-RoFormer~\cite{lu2024bsroformer}, a model typically used for stem separation; the top-ranking submission, xlancelab~\cite{zhu2026xlancelab}, used a BS-RoFormer cascade for denoising, stem separation, and vocal dereverberation. More recently, DTT-BSR+~\cite{ni2026dttbsrplus}, a two-stage generative-regression cascade, reported higher reconstruction fidelity than xlancelab on five of eight stems on its private evaluation split. For restoration specifically, Apollo~\cite{li2025apollo} combines a band-split module with RoFormer- and TCN-based sequence modeling and a GAN-based objective for 44.1\,kHz audio restoration.

The success of these systems also points to an important tension in MSR: separation and restoration are not quite the same problem. Masking-based models such as BS-RoFormer are highly effective at assigning mixture content to the correct source and suppressing interference, but restoration asks for something different: recovering information that may have been altered or lost through nonlinear processing and degradation. A spectral mask can reweight what is present in the mixture, but it cannot directly reconstruct missing timbral detail, transient structure, or spatial cues. This limitation was already observed in the MSRBench baselines, where complex-domain separation models remained insufficient for restoration, motivating restoration-specific and instrument-specific strategies~\cite{zang2025msrbench}. 
We therefore separate these roles in our system: a masking-based model handles source separation, while restoration models directly predict the target spectrum and use an adversarial objective to encourage the reconstruction of plausible fine-grained detail, with source-specific experts further refining each separated stem.

Building on this, we propose a three-stage framework organized around modular specialization: (i) mixture restoration addresses mixture-level degradation before separation; (ii) a single model separates the restored mixture into eight target stems (\textit{vocals}, \textit{guitar}, \textit{keyboard}, \textit{synthesizer}, \textit{bass}, \textit{drums}, \textit{percussion}, \textit{orchestra}) plus an auxiliary residual stem; and (iii) source-specific restoration experts, fine-tuned on real separator artifacts, refine each separated stem individually. Mixture restoration and separation operate globally across all sources, since both act before per-stem structure is introduced, while stem restoration is applied locally, targeting the artifact statistics specific to each source.

\clearpage
Our contributions are three-fold: i) We introduce a three-stage MSR architecture and training regime that pairs separate restoration models at both the mixture and stem level, with a single BSRoformer style multi-stem separator; ii) we measure the contribution of each of the three stages to overall restoration quality; and iii) we consolidate and formalize the stem processing, mixing, mastering, and degradation structure described across \cite{zang2025msr} and \cite{zang2025msrbench} into a single, precise problem formulation.

\section{Method}
\label{sec:method}

\subsection{Structured Problem Formulation}
\label{sec:formulation}

Following \cite{zang2025msr}, we model a mixture as a sum of individually degraded stems. Let $s_1, \dots, s_n$ denote the clean, unprocessed stems for $n$ target instrument classes ($n=8$
in our case),
each typically stereo. Each stem is subjected to a stem-level degradation $f_i \in \mathcal{F}$ (e.g. eq, compression, saturation, reverb).
We model the combination of stems as a sum, though in practice stems may be combined via gain-staging and bus-level processing, which we do not explicitly model:

\begin{equation}
\label{eq:sum}
x = \sum_{i=1}^{n} f_i(s_i).
\end{equation}

Beyond stem mixing, mastering ~\cite{shelvock2012audio} (e.g. broadband and multiband equalization, compression, and limiting) is applied to commercially released music. We model it via a function $m \in \mathcal{M}$, yielding the mastered mixture $\bar{x} = m(x)$. Depending on the distribution channel, $\bar{x}$ may undergo degradation $d \in \mathcal{D}$~\cite{zang2025msrbench}, such as acoustic effects, or codec compression~\cite{sripada2006mp3codec}. 
We write $g := d \circ m \in \mathcal{G}$ for this composite, yielding the released recording $y$, also typically stereo:

\begin{equation}
y = g(x) = d\left(m\left(\sum_{i=1}^{n} f_i(s_i)\right)\right).
\end{equation}

Given only $y$, MSR aims to recover estimates $\hat{s}_1, \dots, \hat{s}_n$ approximating the original stems, requiring inversion of $g$, source attribution, and inversion of the residual $\{f_i\}$ entangled within each source.

\subsection{Proposed Framework}
\label{sec:proposed_framework}

\begin{figure}[t]
    \centering
    \includegraphics[width=\linewidth]{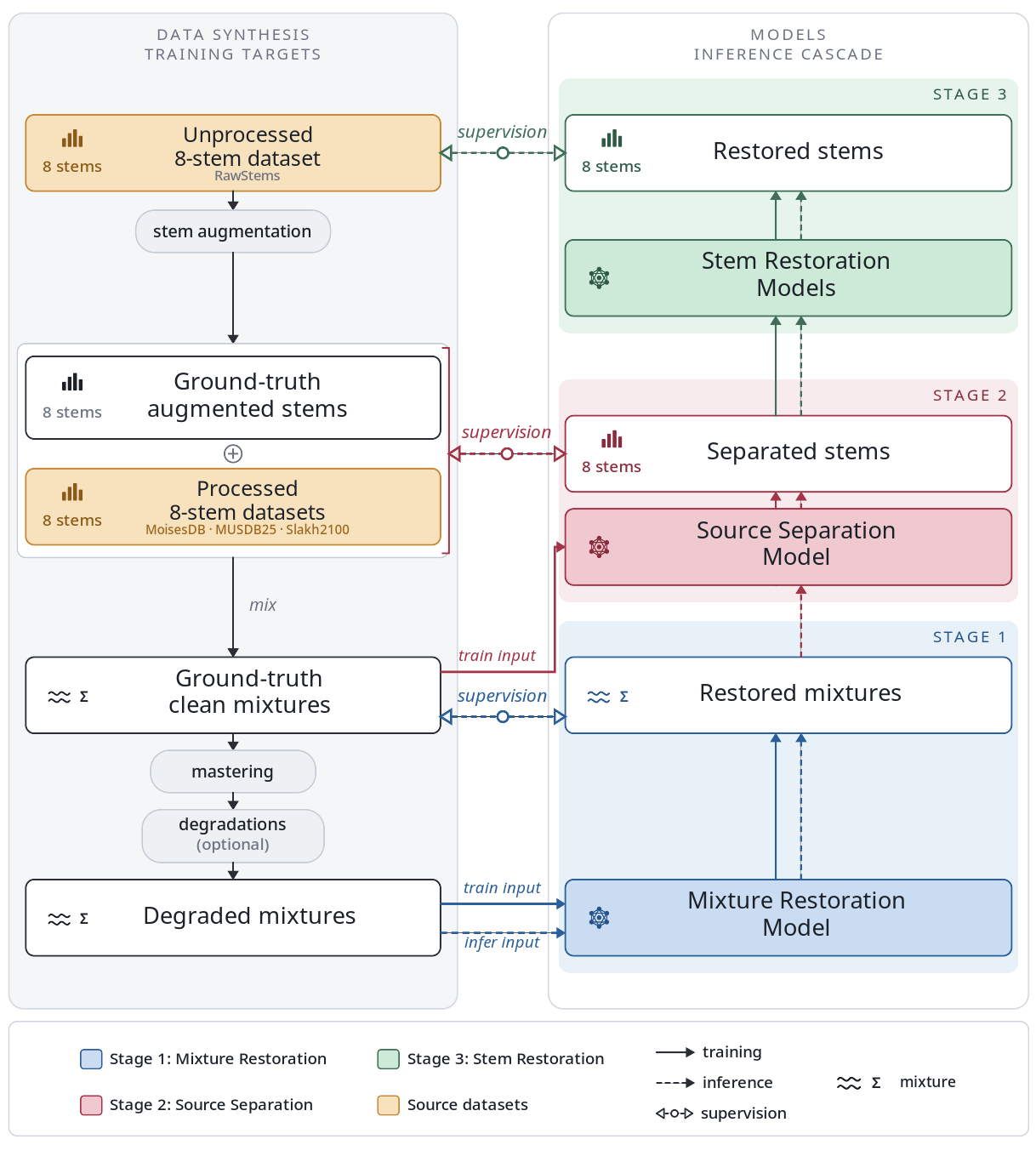}
    \caption{Proposed 3-stage Music Source Restoration pipeline.}
    \label{fig:system_overview}
\end{figure}

Guided by our formulation, our framework aims to recover $\hat{s}_1, \dots, \hat{s}_n$ from a possibly degraded but mastered recording $y$ through three sequential stages, illustrated in Fig.~\ref{fig:system_overview}.

\textbf{Stage 1: Mixture Restoration.} A restoration model inverts the degradation $d$, mapping possibly degraded $y$ to a restored mixture $\hat{x}$ approximating the mastered signal:
\begin{equation}
y \;\mapsto\; \hat{x} \approx \bar{x} = m\left(\sum_{i=1}^{n} f_i(s_i)\right).
\end{equation}

\textbf{Stage 2: 8-stem Source Separation.} A single separation model then attributes the restored mixture $\hat{x}$ to its constituent sources $\hat{u}_i$, recovering each stem as it exists within the mixture, i.e. still carrying its stem-level degradation $f_i$:
\begin{equation}
\hat{x} \;\mapsto\; \{\hat{u}_i\}_{i=1}^{n} \approx \{f_i(s_i)\}_{i=1}^{n}.
\end{equation}
The separation model is trained independently of Stage 1, directly on clean, unmastered mixtures $x$; since Stage 1's output approximates the mastered $\bar{x}$ rather than $x$, this introduces a train/inference mismatch that we leave to future work.

\textbf{Stage 3: Stem Restoration.} Source-specific restoration experts then invert the residual stem-level degradation $f_i$, refining each separated stem individually:
\begin{equation}
\hat{u}_i \;\mapsto\; \hat{s}_i \approx s_i, \quad i = 1, \dots, n.
\end{equation}
Each expert is fine-tuned on real separator artifacts---i.e., on actual outputs $\hat{u}_i$ of Stage 2---rather than on synthetically degraded stems, so that the restoration targets the artifact distribution the separator actually produces.

In contrast to prior modular MSR systems that fine-tune separate models per stem~\cite{zhu2026xlancelab}, we fine-tune a single separation model across all 8 stems, potentially trading some per-stem performance for a single model with substantially lower total parameter count and training cost.


\section{Experimental Setup}
\label{sec:experimental_setup}

\subsection{Data}
\label{sec:data}

\textbf{Datasets.} Our training data combines four multi-track datasets: RawStems~\cite{zang2025msr}, MoisesDB~\cite{pereira2023moisesdb}, Slakh2100~\cite{manilow2019slakh}, and MUSDB25-style extensions~\cite{watcharasupat2025musdb25}. RawStems is the only one containing unprocessed stems; we additionally clean a subset of its annotations to resolve known stem-mixture alignment issues.
For validation and testing, we follow the MSR Challenge~\cite{zang2026msrchallenge} protocol to remain comparable with challenge submissions: MSRBench~\cite{zang2025msrbench} 
serves as our validation set, 
while \textit{testing} uses the recently released MSR Challenge test set\footnote{\url{https://hf.co/datasets/MSRChallenge/2025TestSet}}. Both are released at 48\,kHz; since our models operate at 44.1\,kHz (Section~\ref{sec:models_training}), we resample all validation and test audio accordingly. Unlike a standard validation/test split drawn from one shared pool, MSRBench and the challenge test set are two independent data collection efforts, which may introduce distributional mismatch between them.

\textbf{Mixture Construction for Training.} Following~\cite{zang2025msrbench}, we compute per-second activity masks (RMS $> -40$\,dB) for all training audio segments where at least 80\% of seconds are active across at least two stems.
Since RawStems is the only training dataset with unprocessed stems, we apply a stem-level augmentation pipeline exclusively to RawStems samples, approximating the degradation functions $f_i$ from 
Eq.~(\ref{eq:sum})
each with 10\% probability, followed by RMS-matching to preserve original loudness. 
This is intended to align RawStems with the other, already processed training datasets.
Stems are combined into mixtures $x$ by scaling each relative to a randomly chosen reference stem to a target SNR sampled from $(0,10)$\,dB (gain-staging step mentioned in Section~\ref{sec:formulation}). 

\subsection{Models and Training}
\label{sec:models_training}

All models are fine-tuned on 44.1\,kHz stereo audio; the two Apollo-based restoration models operate on 4sec segments, while the separation model operates on 10sec segments.

\textbf{Mixture Restoration.} The mixture restoration model uses the Apollo~\cite{li2025apollo} architecture, initialized from the official pretrained checkpoint. Apollo splits the input spectrogram into frequency bands, models each band jointly across frequency and time, and reconstructs the output spectrogram directly rather than applying a mask to the input. We freeze the first three BSNet blocks and fine-tune the remaining three and the output head for 20 epochs, using Adam~\cite{kingma2015adam} with a learning rate of $1.5\mathrm{e}{-6}$, exponentially decayed, and an effective batch size of 32.
The training target is the undegraded mixture.
With probability 0.5, 
the target and input both undergo our mastering simulation. 
The input is then further subjected to a degradation such as analog/acoustic effects or codec compression. 


\textbf{Source Separation.}
For separation, we use a pretrained 6-stem BS-RoFormer checkpoint\footnote{\href{https://huggingface.co/jarredou/BS-ROFO-SW-Fixed}{BS-ROFO-SW-Fixed}, downloaded from Hugging Face.} 
and extend it to 9 stems, resulting in 248M parameters. The additional stem is an auxiliary \textit{other} stem, which has effectively no use and whose gradients are not back-propagated during training. Mask-estimation heads for the pretrained stems (\textit{vocals}, \textit{bass}, \textit{drums}, \textit{guitar}, \textit{keyboards}) are initialized from the checkpoint; heads for the new stems (\textit{percussion}, \textit{synthesizers}, \textit{orchestral}) are randomly initialized. We fine-tune the transformer blocks and all mask-estimation heads for about 17{,}000 training steps while keeping the band-split module frozen, using Adam with a linear schedule (8{,}000 warmup steps, peak LR $4\mathrm{e}{-4}$, final LR $1\mathrm{e}{-4}$) and an effective batch size of 32.
We additionally apply random mixing~\cite{jeon2024cacophony}, combining stems drawn from independent songs into synthetic, cacophony-style training mixtures that deliberately violate the beat and tonal coherence of real music. 


\textbf{Stem Restoration.} For each of the eight target stems, we train a separate Apollo model initialized from the fine-tuned mixture restoration model with the same layers frozen. Unlike mixture restoration, training does not use synthetic degradations; instead, each training pair consists of the Stage 2 separator's estimate of a stem---obtained by running it on mixtures already processed by the restoration model---and the corresponding clean stem, generated once from the RawStems subset prior to training. This targets the artifact distribution the separator actually produces, rather than a synthetically degraded approximation of it. We fine-tune for 40 epochs using Adam, exponentially decayed, and an effective batch size of 16, holding out 20\% of each stem's pairs for validation. 
A stem's restoration model is included only if it improves MMSNR over solely separated stems on the validation set.

\textbf{Losses.} The separation model is trained with a multi-resolution STFT loss and an L1 loss, weighted equally. Both restoration models are trained with a multi-resolution mel loss (weight 0.5) and a multi-mel-SNR loss (weight 0.04) that directly optimizes the MMSNR metric (Section~\ref{sec:setup_metrics}), in combination with an adversarial loss. We use a GAN-loss weight of 0.5 for stem restoration and 0.2 for mixture restoration, as we found training at higher weights unstable for the latter.


\subsection{Evaluation Metrics}
\label{sec:setup_metrics}

Following the MSR Challenge protocol~\cite{zang2026msrchallenge}, we report three objective metrics, each comparing restored output against the corresponding clean reference stem: \textbf{Multi-Mel-SNR (MMSNR)} averages a scale-invariant SNR across multiple Mel-spectrogram resolutions. 
We treat MMSNR as our primary metric, as it is the headline metric used for ranking in the MSR Challenge.
In addition, we report \textbf{Zimtohrli}~\cite{alakuijala2025zimtohrli}, which maps both signals through stages of the human auditory pathway before comparing the resulting perceptual representations via windowed structural similarity, and \textbf{FAD-CLAP}~\cite{kilgour2019fadclap} captures distributional and semantic similarity rather than sample-level fidelity.
We report only objective metrics, as no subjective evaluation was conducted for this work. We note, however, that in the MSR Challenge, system rankings under these objective metrics correlated strongly with subjective Mean Opinion Score ratings collected on a separate blind test set (Spearman $\rho=0.9$)~\cite{zang2026msrchallenge}, lending support to their use as a proxy for perceptual restoration quality in this setting.

\section{Result and Analysis}
\label{sec:results}

\textbf{Effect of random mixing.} Preliminary experiments with several fixed mixing probabilities confirmed that random mixing~\cite{jeon2024cacophony} consistently improves separation performance over not applying it at all. Annealing the mixing probability linearly from 0.9 to 0.0 over the first 85\% of training outperformed every fixed probability we evaluated, improving test-set MMSNR from 
3.81 to 3.94 dB, 
averaged over three seeds.

\textbf{Separation performance.} Table~\ref{tab:mixture_restoration_ablation} reports performance of the separation stage on the test set. A single model jointly separating all eight stems reaches an MMSNR of 3.98 dB, a major improvement over the BSRNN baseline reported on MSRBench~\cite{zang2025msrbench} (Table~\ref{tab:compare_to_other_systems_per_source}), despite not relying on the per-stem or per-group specialization used by prior cascaded systems. 

\begin{table}[t] 
\centering 
\caption{Effect of mixture restoration (Stage 1) on separation performance (Stage 2).} 
\label{tab:mixture_restoration_ablation}
\small
\setlength{\tabcolsep}{10pt}
\resizebox{0.8\columnwidth}{!}{%
\begin{tabular}{@{}l ccc@{}}
\toprule
\textbf{Config.} & MMSNR\(\uparrow\) & Zimt\(\downarrow\) & FAD\(\downarrow\) \\
\midrule
Stage 2
  & 3.980 & 0.015 & 0.300 \\
Stages 1+2
  & \textbf{4.047} & \textbf{0.014} & \textbf{0.293} \\
\bottomrule
\end{tabular}%
}
\end{table}

\textbf{Effect of mixture restoration.}
Preconditioning the mixture with Stage 1 before separation further improves all three metrics (see Table~\ref{tab:mixture_restoration_ablation}), most notably MMSNR (3.980$\to$4.047 dB), confirming that removing input-side degradation benefits downstream separation even though the separator itself was never trained on degraded mixtures.

\textbf{Effect of stem restoration.} Table~\ref{tab:stage3_ablation} shows the per-source effect of adding Stage 3 on top of Stages 1+2. For our final system (see Table~\ref{tab:compare_to_other_systems_per_source}), stem restoration is retained per stem only if it improves MMSNR on the MSRBench validation set; on validation, this held for five stems (\textit{vocals}, \textit{guitars}, \textit{synthesizers}, \textit{percussion}, \textit{orchestral}). On the test set, however, only three of these (\textit{vocals}, \textit{guitars}, \textit{synthesizers}) still show an MMSNR improvement, while overall test-set MMSNR increases slightly (4.047$\to$4.050 dB). The three metrics further disagree in aggregate: while overall MMSNR improves, FAD-CLAP degrades (0.274$\to$0.279), and Zimtohrli remains essentially unchanged. This is consistent with the validation--test distribution shift discussed in Section~\ref{sec:data}, and illustrates concretely why validation-based selection under one metric does not guarantee consistent test-set improvement, even across metrics intended to capture related notions of restoration quality.

\begin{table}[t] 
\centering
\caption{Per-source effect of adding stem restoration (Stage 3) on top of mixture restoration and separation (Stages 1+2).}  
\label{tab:stage3_ablation}
\small
\setlength{\tabcolsep}{6pt}
\begin{tabular}{@{}l cc cc cc@{}}
\toprule
 & \multicolumn{2}{c}{MMSNR $\uparrow$}
 & \multicolumn{2}{c}{ZIMT $\downarrow$}
 & \multicolumn{2}{c}{FAD $\downarrow$} \\
\cmidrule(lr){2-3}\cmidrule(lr){4-5}\cmidrule(lr){6-7}
\textbf{Source} & 1+2 & 1+2+3 & 1+2 & 1+2+3 & 1+2 & 1+2+3 \\
\midrule
bass          & \textbf{6.956} & 6.430 & \textbf{0.010} & \textbf{0.010} & 0.391 & \textbf{0.339} \\
drums         & \textbf{5.604} & 5.550 & 0.015 & \textbf{0.014} & 0.399 & \textbf{0.368} \\
vocals        & 1.486 & \textbf{1.535} & \textbf{0.017} & \textbf{0.017} & \textbf{0.295} & 0.306 \\
guitars       & 3.045 & \textbf{3.070} & \textbf{0.013} & \textbf{0.013} & 0.226 & \textbf{0.222} \\
keyboards     & \textbf{6.660} & 6.618 & 0.011 & \textbf{0.010} & \textbf{0.325} & 0.339 \\
synthesizers  & 3.102 & \textbf{3.131} & \textbf{0.016} & \textbf{0.016} & 0.583 & \textbf{0.541} \\
percussions   & \textbf{0.823} & 0.783 & \textbf{0.016} & \textbf{0.016} & 0.878 & \textbf{0.785} \\
orchestral    & \textbf{4.698} & 4.662 & \textbf{0.015} & \textbf{0.015} & \textbf{0.409} & 0.433 \\
\midrule
\textbf{Overall} & \textbf{4.047} & 3.972 & \textbf{0.014} & \textbf{0.014} & 0.293 & \textbf{0.274} \\
\bottomrule
\end{tabular}
\end{table}

\textbf{Comparison to prior systems.} Table~\ref{tab:compare_to_other_systems_per_source} compares our full system against the MSRBench BSRNN baseline~\cite{zang2025msrbench} and the xlancelab system~\cite{zhu2026xlancelab}. Our system improves substantially over the baseline across all stems and metrics, while trailing xlancelab's cascade of dedicated, per-group specialized models on most stems. Notably, our single-model approach exceeds xlancelab's MMSNR on synthesizers (3.131 vs.\ 2.244 dB) and remains close on vocals and drums, suggesting that joint multi-stem modeling is competitive for some stems despite lacking per-stem specialization.

\begin{table}[t] 
\centering
\caption{Per-source comparison of our final system against the BSRNN baseline~\cite{zang2025msrbench} and the xlancelab pipeline~\cite{zhu2026xlancelab}.}  
\label{tab:compare_to_other_systems_per_source}
\footnotesize
\setlength{\tabcolsep}{3pt}
\resizebox{\columnwidth}{!}{%
\begin{tabular}{@{}l ccc ccc ccc@{}}
\toprule
 & \multicolumn{3}{c}{MMSNR\(\uparrow\)}
 & \multicolumn{3}{c}{ZIMT\(\downarrow\)}
 & \multicolumn{3}{c}{FAD\(\downarrow\)} \\
\cmidrule(lr){2-4}\cmidrule(lr){5-7}\cmidrule(lr){8-10}
\textbf{Source} & BL & xlance & Ours & BL & xlance & Ours & BL & xlance & Ours \\
\midrule
bass         & 2.109 & \textbf{8.145} & 6.956 & 0.014 & \textbf{0.009} & 0.010 & 0.641 & \textbf{0.307} & 0.391 \\
drums        & 2.193 & \textbf{5.806} & 5.604 & 0.020 & \textbf{0.014} & 0.015 & 0.568 & \textbf{0.314} & 0.399 \\
vocals       & 1.057 & \textbf{1.596} & 1.535 & 0.019 & \textbf{0.016} & 0.017 & 0.349 & \textbf{0.266} & 0.306 \\
guitars      & 0.994 & \textbf{3.726} & 3.070 & 0.017 & \textbf{0.011} & 0.013 & 0.370 & \textbf{0.191} & 0.222 \\
keyboards    & 1.693 & \textbf{6.882} & 6.660 & 0.017 & 0.012 & \textbf{0.011} & 0.698 & \textbf{0.317} & 0.325 \\
synthesizers & 0.730 & 2.244 & \textbf{3.131} & 0.020 & 0.017 & \textbf{0.016} & 0.665 & \textbf{0.508} & 0.541 \\
percussions  & 0.000 & \textbf{1.200} & 0.783 & 0.032 & 0.018 & \textbf{0.016} & 1.297 & \textbf{0.543} & 0.785 \\
orchestral   & 0.951 & \textbf{6.071} & 4.662 & 0.020 & \textbf{0.014} & 0.015 & 0.521 & \textbf{0.301} & 0.433 \\
\midrule
\textbf{Average} & 1.216 & \textbf{4.459} & 4.050 & 0.020 & \textbf{0.014} & \textbf{0.014} & 0.436 & \textbf{0.210} & 0.279 \\
\bottomrule
\end{tabular}%
}
\end{table}


\section{Conclusion}
\label{sec:conclusion}




Our results demonstrate that a modular, three-stage decomposition is a sound approach to MSR: mixture restoration improves separation across all three metrics, consistent with our argument that separation and restoration call for distinct model classes. Gains from stem restoration, however, do not always transfer from validation to test, symptomatic of broader limitations in current MSR benchmarking: differences between MSRBench and the challenge test set introduce a validation--test distribution shift, while disagreements among MMSNR, Zimtohrli, and FAD-CLAP caution against treating gains in any single metric as comprehensive improvements. Future benchmarks should explicitly define the target degradation regime, distinguish matched-distribution evaluation from cross-condition robustness, and validate MSR-specific metrics against human listening judgments.

\newpage
\section{Acknowledgments}
This work was supported by the European Research Council under the European Union's Horizon 2020 research and innovation programme (Grant No.\ 101019375, Whither Music?). The LIT AI Lab is supported by the Federal State of Upper Austria. The computational results have been achieved using the Austrian Scientific Computing (ASC) infrastructure.

\bibliographystyle{IEEEbib}
\bibliography{strings,refs}

@string{ismir= "{Proceedings of the International Society for Music Information Retrieval Conference (ISMIR)}"}

@string{iclr = "{Proceedings of the International Conference on Learning Representations (ICLR)}"}

@string{icassp = "{IEEE International Conference on Acoustics, Speech, and Signal Processing (ICASSP)}"}

@inproceedings{zhu2026xlancelab,
  author       = {Jinxuan Zhu and
                  Hao Qiu and
                  Haina Zhu and
                  Jianwei Yu and
                  Kai Yu and
                  Xie Chen},
  title        = {The {SJTU} {X-LANCE} Lab System for {MSR} Challenge 2025},
  booktitle      = icassp,
  year         = {2026}
}

@inproceedings{lu2024bsroformer,
  author       = {Wei Tsung Lu and
                  Ju{-}Chiang Wang and
                  Qiuqiang Kong and
                  Yun{-}Ning Hung},
  title        = {Music Source Separation With Band-Split Rope Transformer},
  booktitle    = icassp,    
  year         = {2024},  
}

@inproceedings{zang2025msrbench,
    title     = {{MSRBench}: A Benchmarking Dataset for Music Source Restoration},
  author    = {Zang, Yongyi and Hai, Jiarui and Ge, Wanying and Dai, Zheqi and Wang, Helin and Mitsufuji, Yuki and Kong, Qiuqiang and Plumbley, Mark D.},
  booktitle = icassp,
  year      = {2026}
}

@inproceedings{zang2026msrchallenge,
  author       = {Yongyi Zang and
                  Jiarui Hai and
                  Wanying Ge and
                  Qiuqiang Kong and
                  Zheqi Dai and
                  Helin Wang and
                  Yuki Mitsufuji and
                  Mark D. Plumbley},
  title        = {Summary of The Inaugural Music Source Restoration Challenge},
  booktitle=icassp,
  year         = {2026},

}

@inproceedings{zang2025msr,
  author       = {Yongyi Zang and
                  Zheqi Dai and
                  Mark D. Plumbley and
                  Qiuqiang Kong},
  title        = {Music Source Restoration},
  booktitle    = {{IEEE} International Workshop on Multimedia Signal Processing ({MMSP})},
  year         = {2025}
}

@inproceedings{ni2026dttbsrplus,
  title     = {{DTT-BSR+}: A Generative-Regression Cascade for Music Source Restoration},
  author    = {Ni, Youran and Tan, Shihong and Wang, Yuzhu and Huang, Gongping},
  booktitle = {Proceedings of Interspeech 2026},
  year      = {2026}
}

@inproceedings{li2025apollo,
  author       = {Kai Li and
                  Yi Luo},
  title        = {Apollo: Band-sequence Modeling for High-Quality Audio Restoration},
  booktitle    = icassp,
  year         = {2025}
}

@misc{alakuijala2025zimtohrli,
      title={Zimtohrli: An Efficient Psychoacoustic Audio Similarity Metric}, 
      author={Jyrki Alakuijala and Martin Bruse and Sami Boukortt and Jozef Marus Coldenhoff and Milos Cernak},
      year={2025},
      eprint={2509.26133},
      archivePrefix={arXiv}
}

@inproceedings{kilgour2019fadclap,
  author       = {Kevin Kilgour and
                  Mauricio Zuluaga and
                  Dominik Roblek and
                  Matthew Sharifi},
  title        = {Fr{\'{e}}chet Audio Distance: {A} Reference-Free Metric for Evaluating
                  Music Enhancement Algorithms},
  booktitle    = {{Conference of the International Speech Communication Association}},  
  year         = {2019}
}

@article{cano2019mss,
  author       = {Estefan{\'{\i}}a Cano and
                  Derry FitzGerald and
                  Antoine Liutkus and
                  Mark D. Plumbley and
                  Fabian{-}Robert St{\"{o}}ter},
  title        = {Musical Source Separation: An Introduction},
  journal      = {{IEEE} Signal Process. Mag.},
  volume       = {36},  
  year         = {2019}
}

@mastersthesis{sripada2006mp3codec,
  title  = {{MP3 Decoder in Theory and Practice}},
  author = {Sripada, Praveen},
  school = {Blekinge Institute of Technology},
  year   = {2006},  
  address = {Sweden}
}

@mastersthesis{shelvock2012audio,
  author     = {Shelvock, Matt},
  title      = {Audio Mastering as Musical Practice},
  school     = {The University of Western Ontario},
  year       = {2012},
  type       = {Master's thesis},
  address    = {Canada},
}

@misc{watcharasupat2025musdb25,
  author       = {Watcharasupat, Karn},
  title        = {{MUSDB25}: A Fully Multitrack Dataset for Music Source Separation},
  year         = {2025},
  howpublished = {Hugging Face}, 
  url          = {https://huggingface.co/datasets/kwatcharasupat/musdb25}
}

@inproceedings{pereira2023moisesdb,
  title     = {{MoisesDB}: A Dataset for Source Separation Beyond 4-Stems},
  author    = {Pereira, Igor and Ara{\'u}jo, Felipe and Korzeniowski, Filip and Vogl, Richard},
  booktitle = {Proceedings of the 24th International Society for Music Information Retrieval Conference (ISMIR)},
  pages     = {619--626},
  address   = {Milan, Italy},
  year      = {2023}
}

@inproceedings{manilow2019slakh,
  title={Cutting Music Source Separation Some {Slakh}: A Dataset to Study the Impact of Training Data Quality and Quantity},
  author={Manilow, Ethan and Wichern, Gordon and Seetharaman, Prem and Le Roux, Jonathan},
  booktitle={{Proc. IEEE Workshop on Applications of Signal Processing to Audio and Acoustics (WASPAA)}},
  year={2019}  
}

@inproceedings{jeon2024cacophony,
  author       = {Chang{-}Bin Jeon and
                  Gordon Wichern and
                  Fran{\c{c}}ois G. Germain and
                  Jonathan Le Roux},
  title        = {Why Does Music Source Separation Benefit from Cacophony?},
  booktitle    = icassp,  
  year         = {2024}
}

@inproceedings{kingma2015adam,
  author       = {Diederik P. Kingma and
                  Jimmy Ba},
  title        = {Adam: {A} Method for Stochastic Optimization},
  booktitle    = iclr,
  year         = {2015},  
}

\end{document}